# STABILIZATION AT UPRIGHT EQUILIBRIUM POSITION OF A DOUBLE INVERTED PENDULUM WITH UNCONSTRAINED BAT OPTIMIZATION

Kavirayani Srikanth[1] and G V Nagesh Kumar[2]

[1]Department of Electrical and Electronics Engineering, GVPCE, Visakhapatnam, India
[2]Dept.of EEE,GITAM University,Visakhapatnam, India

## ABSTRACT

*A double inverted pendulum plant has been in the domain of control researchers as an established model for studies on stability. The stability of such as a system taking the linearized plant dynamics has yielded satisfactory results by many researchers using classical control techniques. The established model that is analyzed as part of this work was tested under the influence of time delay, where the controller was fine tuned using a BAT algorithm taking into considering the fitness function of square of error. This proposed method gave results which were better when compared without time delay wherein the calculated values indicated the issues when incorporating time delay.*

## 1.INTRODUCTION

Double Inverted Pendulum(DIP) is a typical underactuated non-linear plant which has potential applications in the field of defense, aerospace, mechatronic systems and other industrial applications which use various levels of manipulators. The plant model has been analyzed for modifications in dynamics and also for extending the model to suit to industry requirements by various researchers. DIP as a system can be used for understanding and aids control education. Various types of controllers such as LQR, LQG, H-Infinity , Kalman filter based observer design have been applied by researchers.

A typical double inverted pendulum has a cart and also two pendulums at the center which are free to oscillate about its axes from the unstable equilibrium position to the stable equilibrium position. Typically the arrangement has a motor driven control for such a system using which the stability can be achieved.

Researchers have developed simulation prototypes for the double inverted pendulum and also have developed toolboxes using which the characteristics of the plant can be obtained by defining parameters such as mass of the double pendulum, length of the pendulum and other specifications of the motors that can be controlled. Active research has happened on other versions of the inverted pendulum systems such as the triple inverted pendulum and quadruple pendulum systems.

Arora et al[1] have indicated the differences between BAT, Firefly and cuckoo search and have indicated that BAT does not store history as a significant difference between the algorithms. [2], [4] have applied BAT algorithm to various domains such as intrusion detection, thyristor control for optimal capacitor placement. Mahindrakar et al [5] have analyzed acrobats, pendubot systems and other structures with varied dynamics for various types of controllers. [6] has used echolocation for STATSCOM design and [7-10] are examples where BAT was applied in dealing with robotic navigation. [11] elaborated on fuzzy controller design for the rotary inverted pendulum. [12] has identified the predominant suffering due to time delays and [13-14] has identified the application to humanoid robot. [15] is the author's work on rotary inverted pendulums using particle swarm which formed the basis for advanced studies. [16],[17] have taken up the general aspects with time delays.

Researchers have tested many versions of the dynamics of the inverted pendulum using various controllers from simple PID to complex H-infinity design and sliding mode controllers.

This paper investigates the case of including a time delay in the signal transmitted to the motor that control the input of the rotary inverted pendulum. The analysis is also extended to the case where a particle swarm optimization technique minimizes the motion of the states of the pendulum and minimizes their displacement resulting in an enhanced controller design.

## 2. MATHEMATICAL MODELING

### 2.1. System Dynamics With Time Delay

A typical arrangement of the double inverted pendulum dynamics involves Euler-Lagrangian equation as in [3]. The important system parameter state variables include X , $\Theta_1$ and $\Theta_2$ which denote the cart position, the lower pendulum angle with respect to the vertical , $\Theta_2$ is the upper pendulum angle with respect to the pendulum and $\Theta_3$ is the angle with respect to the vertical. F is the force acting on the cart. The pendulum and cart masses are $m_1, m_2$ and M. The non-linear model is described for the two stage inverted pendulum by a set of three nonlinear differential equations described by equations 1, 2 & 3.

$$\frac{d}{dt}(\frac{\partial T}{\partial \dot{q}_i}) - \frac{\partial T}{\partial \dot{q}_i} + \frac{\partial U}{\partial \dot{q}_i} = Q_i$$

(1)

Where
$L = T - U$

*(2)*

$T$: kinetic energy
$U$ : potential energy
$Qi$: generalized forces not taken into account in $T,U$
$q_i$: generalized coordinates
and the generalized coordinates can be taken as

$$q_i = \begin{bmatrix} x_c \\ \theta_1 \\ \theta_3 \end{bmatrix}$$

(3)

$$Q_i = \begin{bmatrix} F \\ 0 \\ 0 \end{bmatrix}$$

(4)

The following assumptions are important in the modeling of double inverted pendulum system dynamic analysis.

1) The system starts in a state of initial conditions with the pendulum position in less than 5 degrees away from its unstable equilibrium position.
2) A small perturbation is applied to the pendulum.
The pendulum is represented as given in figure 1 with two links.

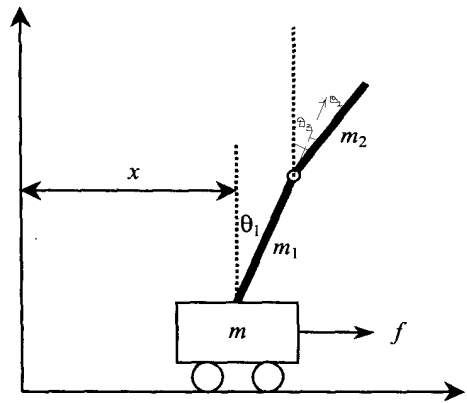

Figure 1. Double Inverted Pendulum

The total kinetic energy for the system is the sum of the kinetic energies for the cart and the two links, is given by equation 5 as

$$T = T_{cart} + T_{pend1} + T_{pend2}$$

(5)

$$T = \frac{1}{2} M \dot{x}_c^2 + \frac{1}{2} m_1 [(\dot{x}_c + l_1 \dot{\theta}_1 \cos\theta_1)^2 + (l_1 \dot{\theta}_1 \sin\theta_1)^2] + \frac{1}{2} J_1 \dot{\theta}_1^2$$
$$+ \frac{1}{2} m_2 [(\dot{x}_c + L_1 \dot{\theta}_1 \cos\theta_1 + l_2 \dot{\theta}_3 \cos\theta_3)^2 + (L_1 \dot{\theta}_1 Sin\theta_1 + l_2 \dot{\theta}_3 \sin\theta_3)^2$$
$$+ \frac{1}{2} J_2 \dot{\theta}_3^2$$

(6)

The total potential energy of the system is the sum of the potential energies of the cart and the two pendulums which is found to be

$$U = U_{Cart} + U_{pend1} + U_{pend2}$$

(7)

$$U = m_1 g l_1 \cos\theta_1 + m_2 g [L_1 \cos\theta_1 + l_2 \cos\theta_3] \qquad (8)$$

Using these equations and based on assumptions as in [3] the system developed would appear as follows:

$$(M + m_1 + m_2)\ddot{x}_C + (m_1 l_1 + m_2 L_1)(\ddot{\theta}_1 \cos\theta_1 - \dot{\theta}_1^2 \sin\theta_1 + m_2 l_2 (\ddot{\theta}_3 \cos\theta_3 - \dot{\theta}_3^2 \sin\theta_3) = F \qquad (9)$$

$$(m_1 l_1 + m_2 L_1)(\ddot{x}_C Cos\theta_1 - gSin\theta_1) + (m_1 l_1^2 + m_2 L_1^2 + J_1)\ddot{\theta}_1$$
$$+ m_2 l_2 L_1 [\ddot{\theta}_3 Cos(\theta_1 - \theta_3) + \dot{\theta}_3^2 Sin(\theta_1 - \theta_3)] = 0 \qquad (10)$$

$$m_2 l_2 (\ddot{x}_c Cos\theta_3 - gSin\theta_3) + (m_2 l_2^2 + J_2)\ddot{\theta}_3$$
$$+ m_2 l_2 L_1 [\ddot{\theta}_1 \cos(\theta_1 - \theta_3) - \dot{\theta}_1^2 \sin(\theta_1 - \theta_3)] = 0 \qquad (11)$$

Where

$$M = M\_C + m_1 + m_2 \qquad (12)$$

A linearized model is obtained by small signal approximation using the above equations. This will yield the state space matrices that define the system for such a model would be given as follows below

$$\begin{bmatrix} \dot{x}_1 \\ \dot{x}_2 \\ \dot{x}_3 \\ \dot{x}_4 \\ \dot{x}_5 \\ \dot{x}_6 \end{bmatrix} = \begin{bmatrix} 0 & 0 & 0 & 1 & 0 & 0 \\ 0 & 0 & 0 & 0 & 1 & 0 \\ 0 & 0 & 0 & 0 & 0 & 1 \\ 0 & A_{42} & A_{43} & A_{44} & 0 & 0 \\ 0 & A_{52} & A_{53} & A_{54} & 0 & 0 \\ 0 & A_{62} & A_{63} & A_{64} & 0 & 0 \end{bmatrix} \begin{bmatrix} x_1 \\ x_2 \\ x_3 \\ x_4 \\ x_5 \\ x_6 \end{bmatrix} + \begin{bmatrix} 0 \\ 0 \\ 0 \\ B_{41} \\ B_{51} \\ B_{61} \end{bmatrix} u \qquad (13)$$

$$\begin{bmatrix} y_1 \\ y_2 \\ y_3 \\ y_4 \\ y_5 \\ y_6 \end{bmatrix} = \begin{bmatrix} 1 & 0 & 0 & 0 & 0 & 0 \\ 0 & 1 & 0 & 0 & 0 & 0 \\ 0 & 0 & 1 & 0 & 0 & 0 \\ 0 & 0 & 0 & 1 & 0 & 0 \\ 0 & 0 & 0 & 0 & 1 & 0 \\ 0 & 0 & 0 & 0 & 0 & 1 \end{bmatrix} \begin{bmatrix} x \\ x_2 \\ x_3 \\ x_4 \\ x_5 \\ x_6 \end{bmatrix} + \begin{bmatrix} 0 \\ 0 \\ 0 \\ 0 \\ 0 \\ 0 \end{bmatrix} u \qquad (14)$$

Let,
$p_1 = (m_1 + 2 * m_2) * L_1;$
$p_2 = m_2 * L_2$

$p_3 = 2 * m_2 * L_1 * L_2$
$p_4 = (m_1 + 4 * m_2) * L_1 * L_2$
$p_5 = m_2 * L_1 * L_1$
$Den = M * p_4 * p_5 + 2 * p_1 * p_2 * p_3 - p_2 * p_2 * p_4 - M * p_3 * p_3 - p_1 * p_1 * p_5$

Where the matrix parameters are

$$A_{42} = \frac{((p_2 * p_3 - p_1 * p_5) * p_1 * g);}{Den}$$

$$A_{43} = \frac{((p_1 * p_3 + p_2 * p_4) * p_2 * g);}{Den}$$

$$A_{44} = \frac{-((p_4 * p_5 - p_3 * p_3) * f);}{Den}$$

$$A_{52} = \frac{((M * p_5 - p_2 * p_2) * p_1 * g)}{Den}$$

$$A_{53} = \frac{-((M * p_3 - p_1 * p_2) * p_2 * g);}{Den}$$

$$A_{54} = \frac{-((p_1 * p_5 - p_2 * p_3) * f}{Den}$$

$$A_{62} = \frac{((M * p_3 - p_1 * p_2) * p_1 * g)}{Den}$$

$$A_{63} = \frac{((M * p_4 - p_1 * p_1) * p_2 * g)}{Den}$$

$$A_{64} = \frac{-((-p_1 * p_3 + p_2 * p_4) * f);}{Den}$$

$$B_{41} = \frac{(p_4 * p_5 - p_3 * p_3)}{Den}$$

$$B_{51} = \frac{(p_1 * p_5 - p_2 * p_3)}{Den}$$

$$B_{61} = \frac{(p_1 * p_3 + p_2 * p_4)}{Den}$$

The states x1 to x7 indicate cart position, cart velocity, upper pendulum position, upper pendulum velocity, lower pendulum position, lower pendulum velocity.

The measurement noise is neglected and only a desired cart position is considered. The time delay parameter which is considered here is at the output measurement which directly affects the system performance. With the help of Pade approximation, the shift operator can be approximated by a rational transfer function of first order as in [12]

$$e^{=s\Gamma} = \frac{(1 - s\Gamma/2)}{(1 + s\Gamma/2)} \tag{15}$$

Where $\Gamma$ in equation refers to the time delay in the control input which is applied to the cart. Now introducing a new state variable into the system as the state x$_7$,

The time delay parameter would dynamics would involve an additional state defined from the following

$$\bar{x}_m(1 + \frac{s\Gamma}{2}) = \bar{x}_1(1 - \frac{s\Gamma}{2}) \tag{16}$$

$$x_m - x_1 = -(\dot{x}_m - \dot{x}_1)\frac{\Gamma}{2} \tag{17}$$

Where $\Gamma$ in equation refers to the time delay in the control input which is applied to the cart position and $X_m$ denotes the modified output state condition. Now introducing a new state variable $X_7$ into the system ,

$$\dot{x}_7 = \frac{2}{\Gamma} x_7 + \frac{-4}{\Gamma} \theta_1 \qquad (18)$$

The modified state matrices because of addition of time delay parameter causes a new state to arise which modifies the system as given below:

The modified state matrices because of addition of time delay parameter causes a new state to arise which modifies the system as given below:

$$\begin{bmatrix} \dot{x}_1 \\ \dot{x}_2 \\ \dot{x}_3 \\ \dot{x}_4 \\ \dot{x}_5 \\ \dot{x}_6 \\ \dot{x}_7 \end{bmatrix} = \begin{bmatrix} 0 & 0 & 0 & 1 & 0 & 0 & 0 \\ 0 & 0 & 0 & 0 & 1 & 0 & 0 \\ 0 & 0 & 0 & 0 & 0 & 1 & 0 \\ 0 & A_{42} & A_{43} & A_{44} & 0 & 0 & 0 \\ 0 & A_{52} & A_{53} & A_{54} & 0 & 0 & 0 \\ 0 & A_{62} & A_{63} & A_{64} & 0 & 0 & 0 \\ \frac{-4}{\Gamma} & 0 & 0 & 0 & 0 & 0 & \frac{2}{\Gamma} \end{bmatrix} \begin{bmatrix} x \\ x_2 \\ x_3 \\ x_4 \\ x_5 \\ x_6 \\ x_7 \end{bmatrix} + \begin{bmatrix} 0 \\ 0 \\ 0 \\ B_{41} \\ B_{51} \\ B_{61} \\ 0 \end{bmatrix} u \qquad (19)$$

$$\begin{bmatrix} y_1 \\ y_2 \\ y_3 \\ y_4 \\ y_5 \\ y_6 \\ y_7 \end{bmatrix} = \begin{bmatrix} 1 & 0 & 0 & 0 & 0 & 0 & 0 \\ 0 & 1 & 0 & 0 & 0 & 0 & 0 \\ 0 & 0 & 1 & 0 & 0 & 0 & 0 \\ 0 & 0 & 0 & 1 & 0 & 0 & 0 \\ 0 & 0 & 0 & 0 & 1 & 0 & 0 \\ 0 & 0 & 0 & 0 & 0 & 1 & 0 \\ 0 & 0 & 0 & 0 & 0 & 0 & 1 \end{bmatrix} \begin{bmatrix} x \\ x_2 \\ x_3 \\ x_4 \\ x_5 \\ x_6 \\ x_7 \end{bmatrix} + \begin{bmatrix} 0 \\ 0 \\ 0 \\ 0 \\ 0 \\ 0 \\ 0 \end{bmatrix} u \qquad (20)$$

## 2.2.BAT Algorithm Methodology

BAT algorithm is a nature inspired algorithm where echolocation behaviour of bats is used as a procedure for obtaining the fitness of the function. The loudness of the shouts and echos of the bat are taken as the parameters of the algorithm and are used for fine tuning the fitness of a function. This algorithm has been extensively used in engineering design optimization and has implications on future research[1], [2], [4], [6]. BATS used only sonar data for detection of food, tracking enemies and locating their surrounding the implementation of an algorithm that mimics the behavior of BAT can be given as follows using the below flow chart as shown in figure 2. Figure 3 indicates the control logic involved for modeling the systems under the influence of time delay.

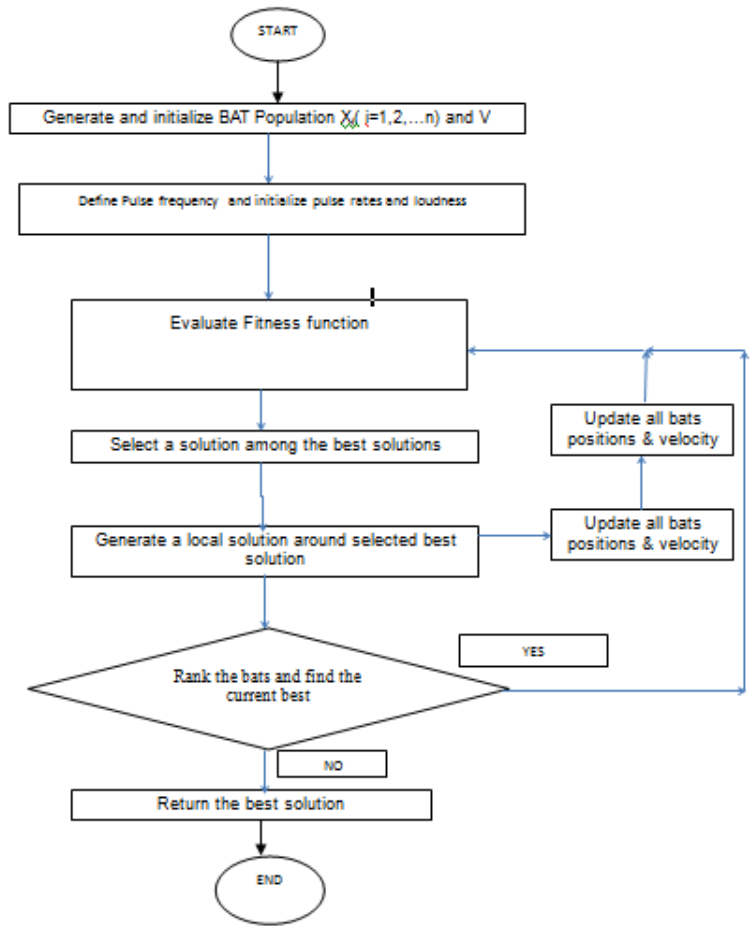

Figure2: Flow chart indicating BAT logic

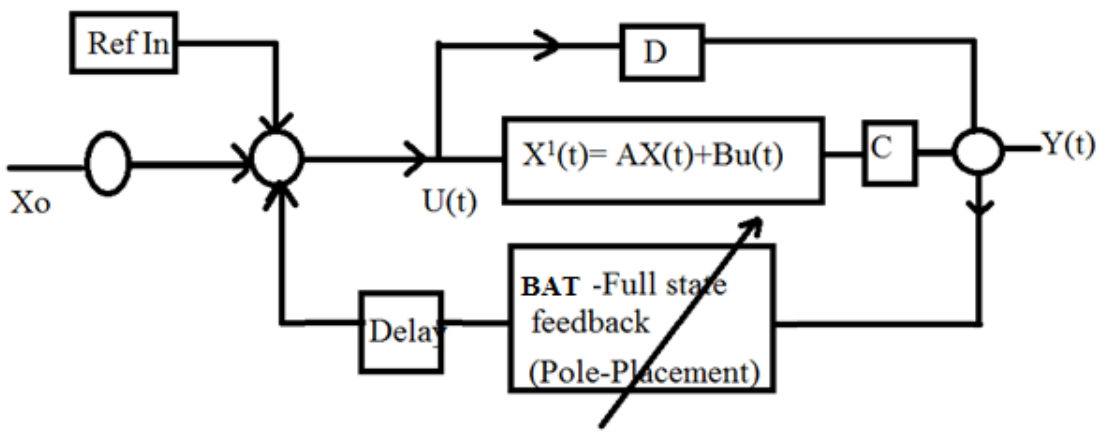

Figure3: Block diagram with BAT based controller and time delay

The bat parameters are given as follows from table II wherein the loudness and pulse rate are selected basically based on various iterations.

TABLE I

| Parameter | Value |
|---|---|
| Population size | 20 |
| Number of generations | 20 |
| Loudness | 0.5 |
| Pulserate | 0.5 |

The loudness and pulse rate determine the speed with which the particular function reaches its optimal value.

## 3.RESULTS

The convergence value of the fitness function is evaluated as follows in figure 4 where within 20 iterations the function has converged to its final value. Figure 4 to 7 indicate the variation of states with and without time delay and it can be clearly seen that control can be achieved if the variation of the parameters of BAT are arranged in such a way that there would be less impact if the time delay is less.

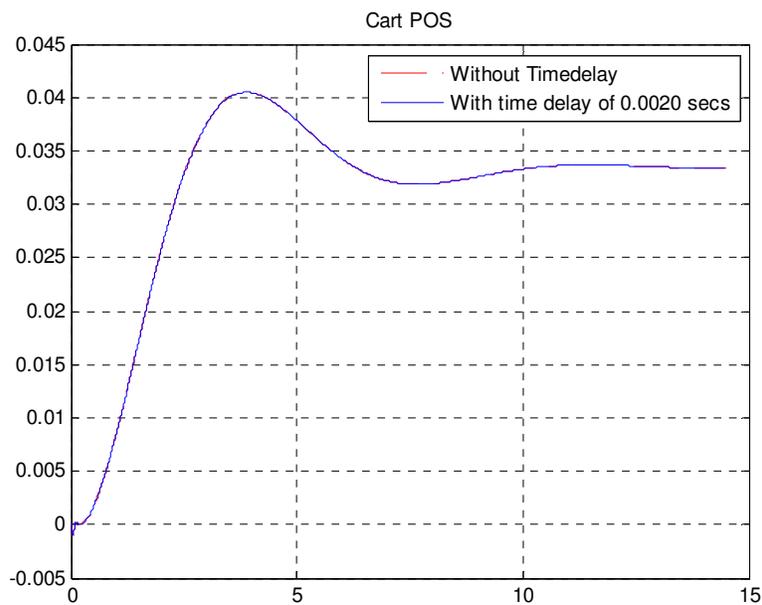

Figure 4: Double Inverted Pendulum cart position

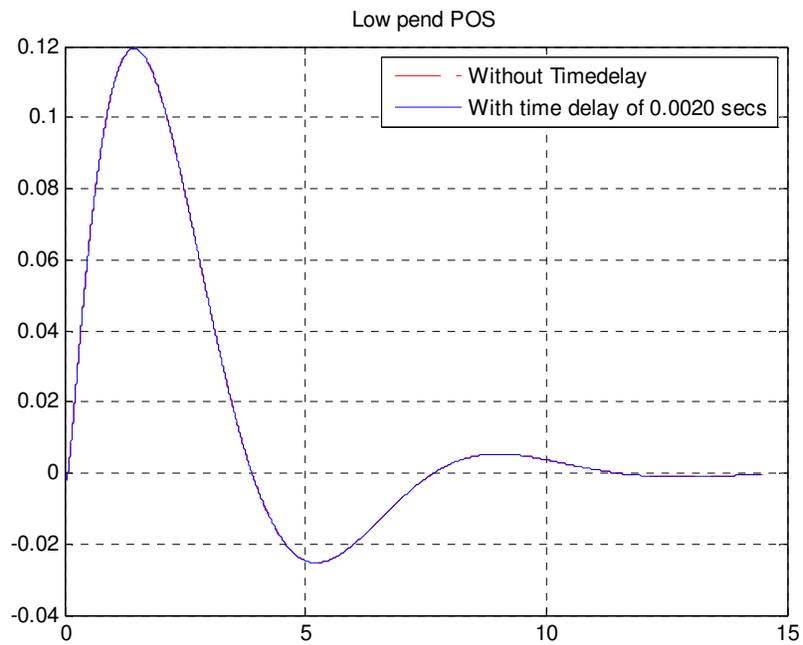

Figure 5: Double Inverted Pendulum lower pendulum position

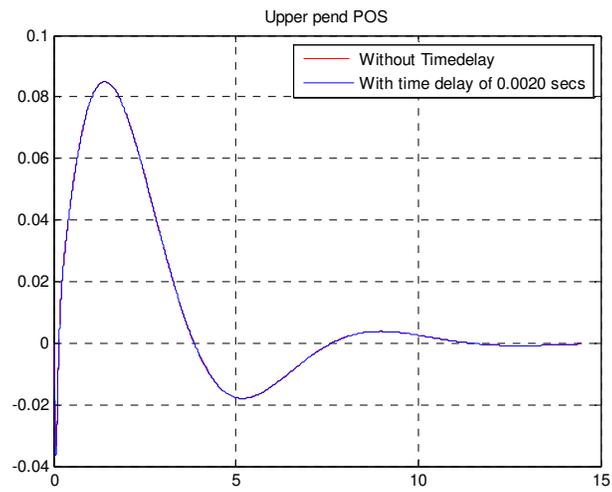

Figure 6: Double Inverted Pendulum Upper pendulum position

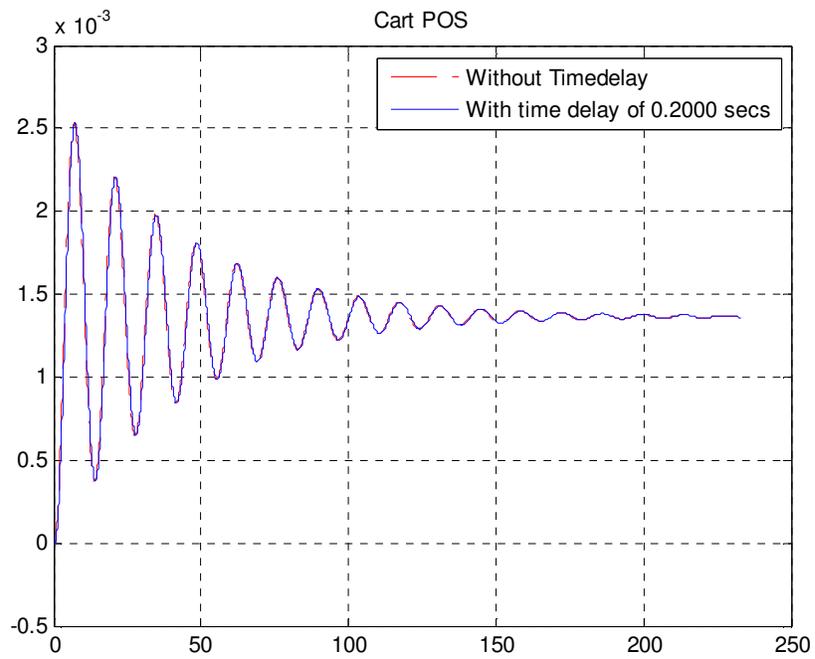

Figure 7: Double Inverted Pendulum cart position

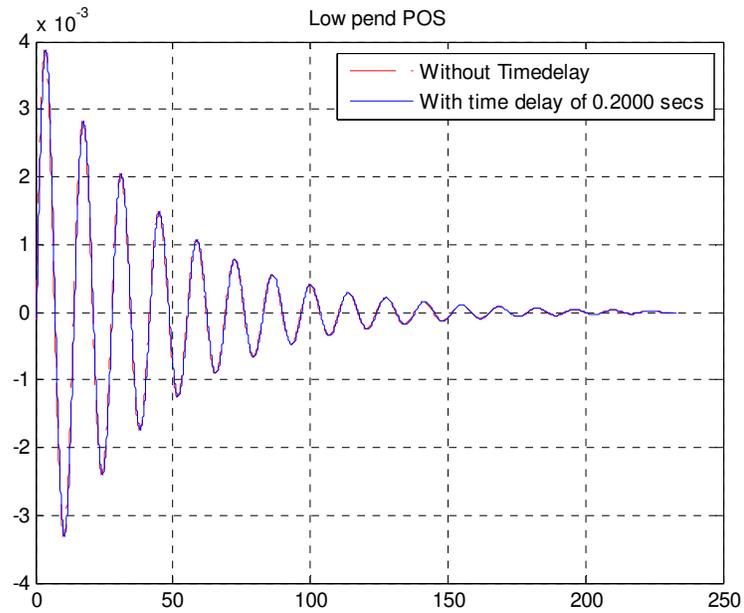

Figure 8: Double Inverted Pendulum lower pendulum position

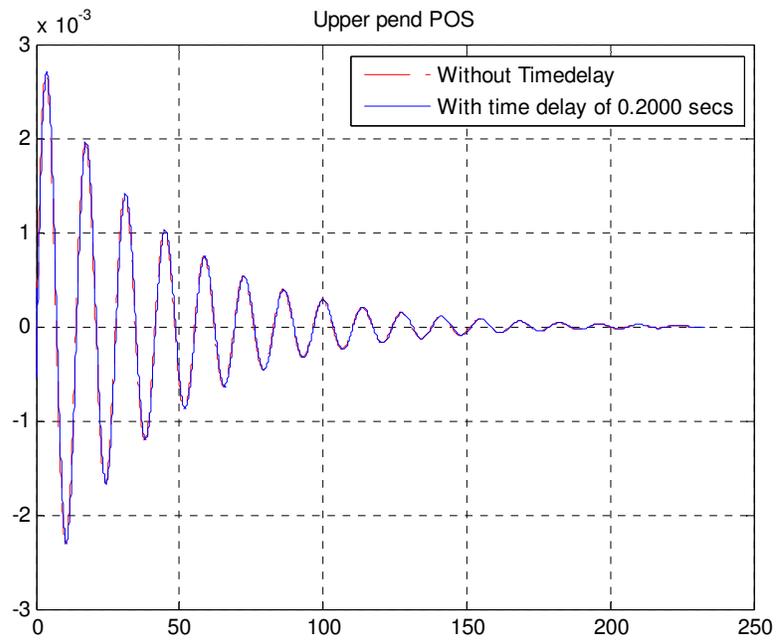

Figure 9: Double Inverted Pendulum upper pendulum position

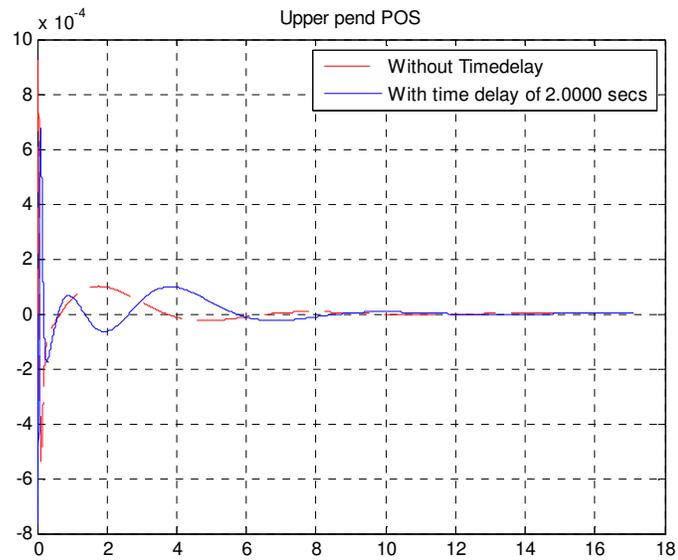

Figure 10: Double Inverted Pendulum upper pendulum position

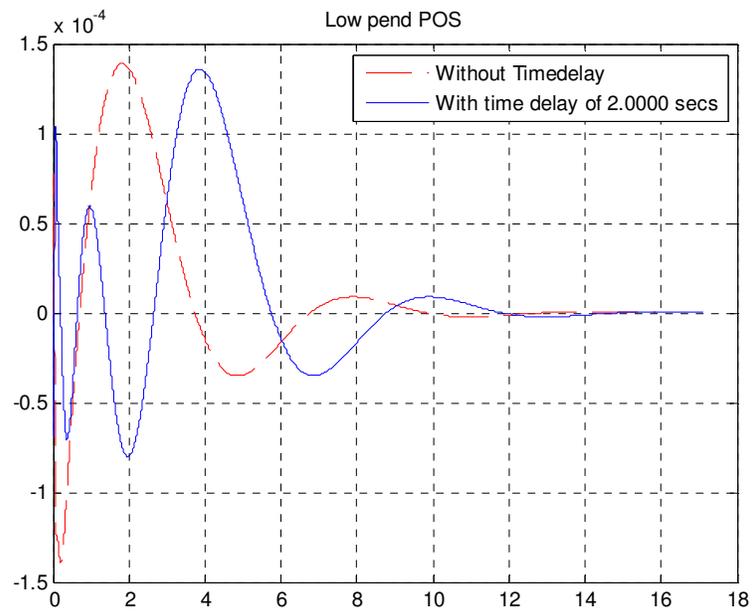

Figure 11: Double Inverted Pendulum lower pendulum position

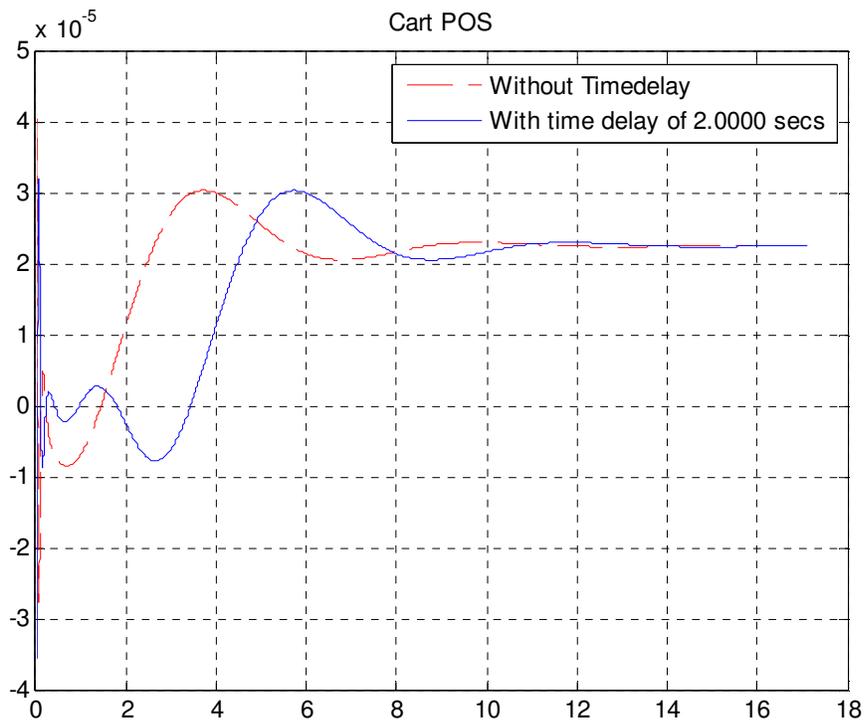

Figure 12: Double Inverted Pendulum cart position

The variation of the pendulum cart position, lower pendulum and upper pendulum can be seen for various time delays. When the time delay was very less in figures 4 to 6, the output response coincided for the system with and without time delays. However as the time delay is increased in figures 7 to 9 , the variation was seen and was significantly seen in figures 10 to 12. In figures 10 to 12, the time delay impact is felt however the system could be stabilized and the time of computation was less for higher time delays and the damping factor and natural frequency were designed optimally by the bat algorithm. The Bat algorithm was used with a set of defined boundaries for both the damping factor and natural frequency as the variables that optimized a fitness of square of the error of input.

TABLE II Double Pendulum Analysis with BAT Algorithm

| S.no | Time delay(Secs) | Time of computation(Secs) | Damping Factor | Natural frequency(rad/s) |
|------|------------------|---------------------------|----------------|--------------------------|
| 1 | 0.02 | 11.073860 | 0.4431 | 0.9248 |
| 2 | 0.2 | 4.03 | 0.0512 | 0.4570 |
| 3 | 2.0 | 3.928 | 0.4013 | 1.1270 |

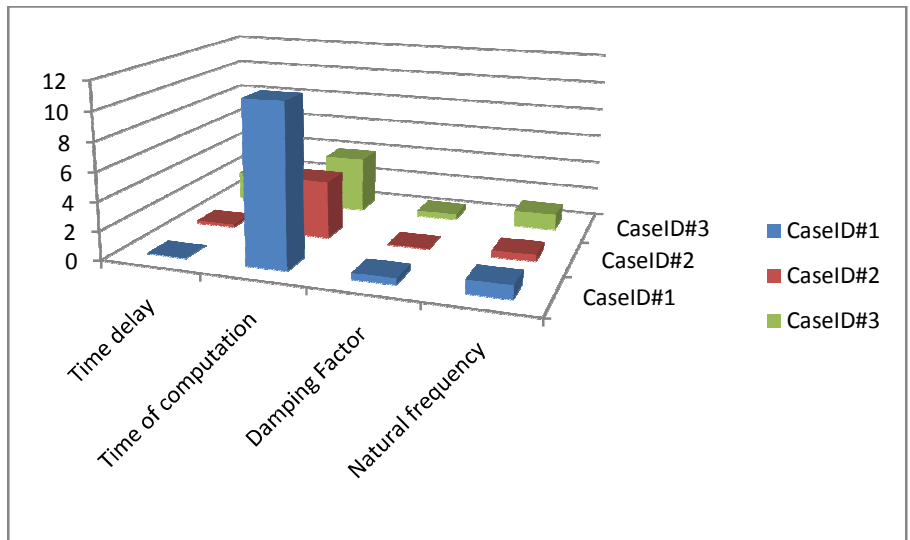

Figure 13: 3-D case study analysis

It can be clearly seen from figure 13 that the case id #1 where the time delay was less had higher computation time and the time of computation has gone down in case id #2 and #3. The natural frequency and damping factor that are considered for the design of the pendulum system stabilization are well in the limits of the system.

## 4.CONCLUSIONS

The system is as such controllable and can be stabilized with a normal LQR controller or a controller designed with full state feedback. In this case it is clearly seen that the system is stabilizable even under the presence of time delay and it is to be seen that the benefit of using bat algorithm is to see that the convergence is achieved at a faster rate for stabilizing the pendulum system.

## ACKNOWLEDGEMENTS

The authors would like to thank his mentors for encouragement and support!

## REFERENCES

[1]   Arora, S.; Singh, S.(2013), "A conceptual comparison of firefly algorithm, bat algorithm and cuckoo search," Control Computing Communication & Materials (ICCCCM), 2013 International Conference on , vol., no., pp.1-4

[2]   Enache, A.-C.; Sgarciu, V., (2014), "Anomaly intrusions detection based on support vector machines with bat algorithm," 2014 18th International Conference System Theory, Control and Computing (ICSTCC) vol., no., pp.856-861

[3]   Srikanth,K.;Nagesh Kumar, G.V.,(2014) "Stabilization at Unstable Equilibrium Position of a Double Inverted Linear Pendulum using Modified Particle Swarm Optimization", Proceedings of  38th National system conference on real time systems –modeling, Analysis and control , 5th -7th Nov, 2014 in JNTU Hyd, Telangana, India. pp.160-164.

[4]   Manikanta, G.; Nagesh Kumar, G.V.,(2014), "Thyristor Controlled Series Capacitor placement and sizing using BAT search optimizer to enhance power flow," 2014 International Conference on Computation of Power, Energy, Information and Communication (ICCPEIC),  vol., no., pp.289-293

[5]   M.T. Ravichandran, A.D. Mahindrakar, (2011), "Robust stabilization of a class of underactuatedmechanicalsystems using time-scaling and Lyapunov redesign", IEEE Transactions on Industrial Electronics, Vol. 58, No.9, pp. 4299-4313.

[6]   Kumaravel, G.; Kumar, C.,(2012) "Design of self tuning PI controller for STATCOM using Bats Echolocation Algorithm based Neural controller Advances in Engineering, Science and Management (ICAESM),", 2012 International Conference on , vol., no., pp.276-281

[7]   Tesheng Hsiao; Chang-Mou Yang; I-Hsi Lee; Chin-Chi Hsiao,(2014), "Design and implementation of a ball-batting robot with optimal batting decision making ability,"2014 IEEE International Conference on Automation Science and Engineering (CASE), vol., no., pp.1026-1031

[8]   Xin-Zhi Zheng; Inamura, W.; Shibata, K.; Ito, K.,(1999) "Task skill formation via motion repetition in robotic dynamic manipulation," 1999. IEEE SMC '99 Conference Proceedings,1999 IEEE International Conference on Systems, Man, and Cybernetics, vol.4, no., pp.1001-1006 vol.4

[9]   Xin-Zhi Zheng; Inamura, W.; Shibata, K.; Ito, K., (2000)"A learning and dynamic pattern generating architecture for skilful robotic baseball batting system," Proceedings. ICRA '00. IEEE International Conference on Robotics and Automation, vol.4, no., pp.3227-3232 vol.4

[10]  Horiuchi, T.K.,(2006) "A neural model for sonar-based navigation in obstacle fields," ISCAS 2006 Proceedings, IEEE International Symposium on Circuits and Systems , vol., no., pp.4

[11]  Sheng Liu; Lihui Cui; Jie Chen; Ming Bai, (2002), "Research of rotary inverted pendulum using fuzzy strategy based on dynamic query table," 2002 Proceedings of the 4th World Congress on Intelligent Control and Automation,, vol.4, no., pp.3161,3165 vol.4

[12]  V. P. Makkapati, M. Reichhartinger and M. Horn, (2013),"The output tracking for dual-inertia system with dead time based on stable system center approach", 2013 IEEE Multi-conference on Systems and Control, Conference on Control Applications, Hyderabad, India, pp.923-928

[13]  Youngbum Jun; Alspach, A.; Oh, P.,(2012) "Controlling and maximizing humanoid robot pushing force through posture,"9th International Conference on Ubiquitous Robots and Ambient Intelligence (URAI), vol., no., pp.158-162.

[14]  Yao Li; Levine, W.S.; Loeb, G.E., (2012)"A Two-Joint Human Posture Control Model With Realistic Neural Delays," IEEE Transactions on Neural Systems and Rehabilitation Engineering, vol.20, no.5, pp.738-748

[15] Srikanth, K.;Nagesh Kumar, G.V., (2014), "Design and Control of Pendulum with Rotary Arm using Full State Feedback Controller based Particle Swarm Optimization for Stabilization of Rotary Inverted Pendulum", Proceedings of 38th National system conference on real time systems – modeling, Analysis and control, JNTU Hyd, Telangana, India. pp.59-63.

[16] Wu, X.-J.; Wu, X.-L.; Luo, X.-Y.; Guan, X.-P., (2012), "Brief Paper - Dynamic surface control for a class of state-constrained non-linear systems with uncertain time delays," Control Theory & Applications, IET , vol.6, no.12, pp.1948-1957

[17] Mazenc, F.; Malisoff, M.; Niculescu, S.-I.,(2014) "Reduction Model Approach for Linear Time-Varying Systems With Delays," IEEE Transactions on Automatic Control, vol.59, no.8, pp.2068-2082

**Authors**

Dr. G.V. Nagesh Kumar was born in Visakhapatnam, India in 1977. He graduated from College of Engineering, Gandhi Institute of Technology and Management, Visakhapatnam, India in 2000, Masters Degree from the College of Engineering, Andhra University, Visakhapatnam, in 2002. He received his Doctoral degree from Jawaharlal Nehru Technological University, Hyderabad in 2008. He has published 92 research papers in national and international conferences and journals. He received "Sastra Award" , "Best Paper Award" and "Best Researcher Award". He is a member of various societies, ISTE, IEEE, IE and System Society of India. He is also a reviewer for IEEE Transactions on Dielectrics and Electrical Insulation, Power Systems and a member on Board of several conferences and journals. 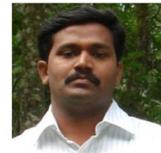

K.Srikanth was born in Visakhapatnam, India. He graduated from gayatriVidyaparishad college of engineering, Visakhapatnam in electrical and electronics engineering in 2002, obtained Master s degree from University of Missouri-Columbia in 2005. He is a part time research scholar in Gitam University and is working as assistant professor in GayatriVidyaParishad College of Engineering, Visakhapatnam, India. 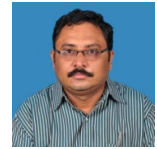